\documentclass[12pt,a4paper]{article}
\usepackage{cite}
\usepackage{fullpage}
\usepackage[centertags]{amsmath}
\allowdisplaybreaks[4]
\usepackage{amssymb}
\usepackage{bm}
\usepackage[dvips]{graphicx}
\usepackage{url}
\usepackage[hyperindex]{hyperref}
\usepackage{color}

\begin{document}

\begin{center}
  {\large\bf Unambiguous Determination of the Neutrino Mass Hierarchy Using Reactor Neutrinos}

\end{center}

\vspace{0.3cm}

\begin{center}
{Yu-Feng Li, Jun Cao, Yifang Wang, Liang Zhan} \\
{ \sl Institute of High Energy Physics, Chinese Academy of Sciences,
Beijing 100049, China}
\end{center}

\setcounter{footnote}{0}

\vspace{2.5cm}

\begin{abstract}
Determination of the neutrino mass hierarchy in a reactor neutrino
experiment at the medium baseline is discussed. Observation of the
interference effects between the $\Delta m^2_{31}$ and $\Delta
m^2_{32}$ oscillations enables a relative measurement independent of
the knowledge of the absolute mass-squared difference. With a 20
kton liquid scintillator detector of the 3\%/$\sqrt{E(\rm MeV)}$
energy resolution, the Daya Bay II Experiment at a baseline of
$\sim$ 50 km from reactors of total thermal power 36 GW can
determine the mass hierarchy at a confidence level of $\Delta
\chi^2_{\text{MH}}\sim(10\div12)$ ($3\div3.5\sigma$) in 6 years
after taking into account the real spatial distribution of reactor
cores. We show that the unknown residual energy non-linearity of the
liquid scintillator detector has limited impact on the sensitivity
due to the self-calibration of small oscillation peaks. Furthermore,
an extra increase of $\Delta \chi^2_{\text{MH}}\simeq4\,(9)$ can be
obtained, by including the precise measurement of the effective
mass-squared difference $\Delta m^2_{\mu\mu}$ of expected relative
error $1.5\%$ ($1\%$) from ongoing long-baseline muon neutrino
disappearance experiments. The sensitivities from the interference
and from absolute measurements can be cross checked. When combining
these two, the mass hierarchy can be determined at a confidence
level of $\Delta \chi^2_{\text{MH}}\sim(15\div20)$ (4$\sigma$) in 6
years.

\end{abstract}

\newpage

\section{Introduction}
An unexpectedly large value of neutrino mixing angle $\theta_{13}$
was measured by the Daya Bay Reactor Neutrino Experiment (DYB)
\cite{DYB,DYBcpc}, together with other consistent evidences from the
reactor \cite{DC,RENO} and accelerator \cite{T2K,MINOS} neutrino
experiments. It opens a gateway to the measurement of neutrino mass
hierarchy (i.e., the sign of $\Delta m^2_{31}$ or $\Delta m^2_{32}$)
and the leptonic CP-violating phase ($\delta_{\rm CP}$). A large
value of $\theta_{13}$ makes both above measurements easier in the
next generation of neutrino oscillation experiments. Possible
information on the neutrino mass hierarchy (MH) can be extracted not
only from the matter-induced oscillations of long-baseline
accelerator neutrino experiments \cite{HK,LBNE,LBNO} and atmospheric
neutrino experiments \cite{INO,PINGU}, but also from the vacuum
oscillation in reactor neutrino experiments at medium baseline
\cite{Reactor1,Reactor2,Reactor3,DYB2t,DYB2e,Reactor4,Reactor5,talk1,talk2,talk3,Emilio,Xin}.

The MH sensitivity of a future reactor neutrino experiment comes
from the interference effect of two separated oscillation modes
\cite{DYB2t,DYB2e} (i.e., two fast oscillations driven by
 $\Delta m^2_{31}$ and $\Delta
m^2_{32}$). The relative sizes of $|\Delta m^2_{31}|$ and $|\Delta
m^2_{32}|$ and the non-maximal value of $\theta_{12}$ make it
possible to determine the MH in vacuum, immune from the uncertainty
of the Earth density profile and ambiguity of the CP-violating phase
in atmospheric and long-baseline accelerator neutrino oscillation
experiments. Such a measurement is challenging. The energy
resolution as the size of $\Delta m^2_{21}/|\Delta m^2_{31}|$ and
event number of several tens of thousands are the minimal
requirements~\cite{DYB2e}. Moreover, the spatial distribution of
reactor cores~\cite{talk1,talk2,talk3} and non-linearity of the
energy response~\cite{Xin} may also degrade the MH sensitivity.

Besides the interference effect in neutrino vacuum oscillations,
direct measurements of the flavor-dependent effective mass-squared
differences (i.e., $\Delta m^2_{ee}$, $\Delta m^2_{\mu\mu}$ and
$\Delta m^2_{\tau\tau}$) \cite{Parke,Kayser} may also include the MH
information. $\Delta m^2_{ee}$ from short base-line reactor neutrino
experiments (i.e., Daya Bay \cite{DYBproposal}) and $\Delta
m^2_{\mu\mu}$ from the long-baseline accelerator muon-neutrino
disappearance are different combinations of $\Delta m^2_{31}$ and
$\Delta m^2_{32}$. Reactor neutrino experiment at the medium
baseline can measure both $\Delta m^2_{31}$ (or $\Delta m^2_{32}$)
and $\Delta m^2_{21}$ up to the MH sign. Therefore, a comparison of
the effective mass-squared differences in different oscillation
scenarios can discriminate the neutrino MH. Considering the MH
determination from the Daya Bay II reactor neutrino Experiment, the
sensitivity from the interference effect can be improved by
including the accurate measurement of the effective mass-squared
difference $\Delta m^2_{\mu\mu}$ from accelerator neutrino
experiments or $\Delta m^2_{ee}$ from reactor neutrino experiments.

Such a study has not been done before and makes sense at least in
the following four aspects. (a) we do the first analysis of the
non-linearity effect with explicit non-linearity functions; (b) we
propose the new idea of non-linearity self-calibration to reduce the
non-linearity effect; (c) our strategy to improve the MH sensitivity
using the effective neutrino mass-squared differences is different
from previous publications; (d) we discuss the effect of baseline
differences and provide the real baseline distribution of the
approved Daya Bay-II experiment.

The outline of this work is planned as follows. We first give a
brief description on the effective mass-squared differences in
different oscillation channels in Section 2. Statistical analysis of
Daya Bay II is introduced in Section 3. Impact from the energy
non-linearity of the liquid scintillator detector is discussed in
Section 4. Improved MH sensitivity with the absolute mass-squared
difference $\Delta m^2_{ee}$ and external $\Delta m^2_{\mu\mu}$
measurements is presented in Section 5. Finally, we conclude in
Section 6.

\section{Effective Mass-Squared Differences}
In the standard three-neutrino mixing scheme the survival
probability for the $\alpha$-flavor neutrinos is given by
\cite{PDG2012}
\begin{eqnarray}
  P(\nu_\alpha \to \nu_\alpha) = P(\bar{\nu}_\alpha \to \bar{\nu}_\alpha) =
  1&-&  4|U_{\alpha3}|^2|U_{\alpha1}|^2 \sin^2 \Delta_{31} \nonumber \\
   &-& 4|U_{\alpha3}|^2|U_{\alpha2}|^2 \sin^2 \Delta_{32}
   \label{alphasurvP} \\
   &-&  4|U_{\alpha2}|^2|U_{\alpha1}|^2 \sin^2 \Delta_{21}, \nonumber
\end{eqnarray}
where $\Delta_{ij} = \Delta m^2_{ij} L /4E$, $\Delta m^2_{ij}=m^2_i
- m^2_j$ and $U_{\alpha i}$ is the element of the leptonic mixing
matrix. Note that only two of the three $\Delta_{ij}$ are
independent because we have the relation $\Delta m^2_{31}=\Delta
m^2_{32}+\Delta m^2_{21}$ from their definitions.

Strong hierarchy (i.e., $\Delta m^2_{21} \ll |\Delta m^2_{31}|$)
between the magnitudes of $\Delta m^2_{21}$ and $\Delta m^2_{31}$
(or $\Delta m^2_{32}$) is achieved from the analysis of solar (or
KamLAND) and atmospheric (or long-baseline accelerator) neutrino
oscillation data \cite{global1,global2,global3}. To separate the
fast and slow oscillation modes, we can rewrite the probability in
Eq. (\ref{alphasurvP}) as
\begin{eqnarray}
  P(\nu_\alpha \to \nu_\alpha) =
  1&-&  4|U_{\alpha3}|^2(1-|U_{\alpha3}|^2) \sin^2 \Delta_{\alpha\alpha} \label{alphasurvPpow}  \\
  &-& 4|U_{\alpha3}|^2 |U_{\alpha1}|^2\sin{[(1-\eta_\alpha)
  \Delta_{21}]}\sin[2\Delta_{\alpha\alpha}+(1-\eta_\alpha)\Delta_{21}] \nonumber \\
  &+& 4|U_{\alpha3}|^2 |U_{\alpha2}|^2\sin{[\eta_\alpha
  \Delta_{21}]}\sin[2\Delta_{\alpha\alpha}-\eta_\alpha\Delta_{21}] \nonumber \\
   &-&  4|U_{\alpha2}|^2|U_{\alpha1}|^2 \sin^2 \Delta_{21},
   \nonumber 
\end{eqnarray}
where an effective mass-squared difference is defined as the linear
combination of $\Delta m^2_{31}$ and $\Delta m^2_{32}$,
\begin{eqnarray}
  \Delta m^2_{\alpha\alpha} &\equiv& \eta_{\alpha} \Delta m^2_{31}+(1-\eta_{\alpha})\Delta
  m^2_{32}\nonumber \\
  &=&\Delta m^2_{32} + \eta_{\alpha}\Delta
  m^2_{21}=\Delta m^2_{31} - (1-\eta_{\alpha})\Delta
  m^2_{21}\,,
  \label{dmeff}
\end{eqnarray}
and $\Delta_{\alpha\alpha}= \Delta m^2_{\alpha\alpha} L /4E$. We can
choose proper values of $\eta_{\alpha}$ to eliminate the terms in
the second and third lines of Eq. (\ref{alphasurvPpow}), and
therefore keep the independent fast and slow oscillation terms. In
general, $\eta_{\alpha}$ is not only the function of neutrino mass
and mixing parameters, but also the function of the neutrino energy
and the baseline.

At the first oscillation maximum of the atmospheric mass-squared
difference, we have $\Delta_{21}\ll 1$ and the corresponding
oscillation effect is extremely small. Expanding to the linear term
of $\Delta_{21}$, we can obtain an effective two-neutrino
oscillation scheme, if the parameter $\eta_{\alpha}$ satisfies the
following relation,
\begin{eqnarray}
    \eta_{\alpha}\simeq\frac{|U_{\alpha1}|^2}{|U_{\alpha1}|^2+|U_{\alpha2}|^2}\,.
    \label{etadm31}
\end{eqnarray}
In a neutrino oscillation experiment of this type, such as the short
base-line reactor neutrino experiment (i.e., Daya Bay) or the
long-baseline accelerator muon-neutrino disappearance experiment, it
is impossible to distinguish between the two neutrino mass
hierarchies because two degenerate solutions with identical $|\Delta
m^2_{\alpha\alpha}|$ but different hierarchies can generate the
identical neutrino energy spectrum. The absolute values of $\Delta
m^2_{31}$ (or $\Delta m^2_{32}$) in the two solutions are different
due to non-zero $\Delta m^2_{21}$. The value of $\eta_{\alpha}$
varies for different oscillation channels due to the
flavor-dependent amplitudes in the oscillation probabilities, so the
degeneracy of the neutrino MH can be removed by comparing the
effective mass-square differences of different neutrino flavors
\cite{Parke,Kayser}.

Using the standard parametrization of the leptonic mixing matrix
\cite{PDG2012}, we get the effective mass-squared differences in
Eq.~(\ref{dmeff}) for different channels of neutrino oscillations
\begin{eqnarray}
\Delta m^2_{ee} &\simeq& \cos^2\theta_{12}\Delta m^2_{31} +
\sin^2\theta_{12}\Delta m^2_{32}\,, \label{dmee}\\
\Delta m^2_{\mu\mu} &\simeq& \sin^2\theta_{12}\Delta m^2_{31} +
\cos^2\theta_{12}\Delta m^2_{32} +
\sin2\theta_{12}\sin\theta_{13}\tan\theta_{23}\cos\delta \Delta m^2_{21}\,,\label{dmmu}\\
\Delta m^2_{\tau\tau} &\simeq& \sin^2\theta_{12}\Delta m^2_{31} +
\cos^2\theta_{12}\Delta m^2_{32} -
\sin2\theta_{12}\sin\theta_{13}\cot\theta_{23}\cos\delta\Delta
m^2_{21}\label{dmtau}\,,
\end{eqnarray}
where terms at the order of ${\cal O}(\sin^2\theta_{13}\Delta
m^2_{21})$ have been neglected for simplicity. We can also
calculate the differences of the effective quantities between
different flavors as
\begin{eqnarray}
|\Delta m^2_{ee}|-|\Delta m^2_{\mu\mu}|&=& \pm \Delta
m^2_{21}(\cos2\theta_{12}-\sin2\theta_{12}\sin\theta_{13}\tan\theta_{23}\cos\delta)\label{dmemu}\,,
\\
|\Delta m^2_{\mu\mu}|-|\Delta m^2_{\tau\tau}|&=& \pm 2\Delta
m^2_{21}
\sin2\theta_{12}\sin\theta_{13}\csc2\theta_{23}\cos\delta\label{dmmutau}\,,
\end{eqnarray}
where the positive and negative signs correspond to normal and
inverted mass hierarchies, respectively.

On the other hand, at the first oscillation maximum of the solar
mass-squared difference, such as the reactor neutrino experiment at
the medium baseline, we have the approximation of
$\sin\Delta_{21}\sim 1$ and $\cos\Delta_{21}\sim 0$. Therefore, we
can separate the fast and slow oscillation terms, if $\eta_{\alpha}$
fulfills the equation as
\begin{eqnarray}
|U_{\alpha1}|^2\cos[\eta_\alpha
  \Delta_{21}]\cos[2\Delta_{32}+\eta_\alpha\Delta_{21}]+
  |U_{\alpha2}|^2\sin[\eta_\alpha
  \Delta_{21}]\sin[2\Delta_{32}+\eta_\alpha\Delta_{21}]=0\,.
  \label{etadm21}
\end{eqnarray}
One should note that $\eta_{\alpha}$ depends on both the neutrino MH
and the neutrino energy. The MH sensitivity is encoded in the energy
dependence of $\Delta m^2_{\alpha\alpha}$. Moreover, because of the
different definitions  of $\Delta m^2_{\alpha\alpha}$ in
these two oscillation scenarios, the MH sensitivity of the reactor
neutrino experiment at the medium baseline can be improved by
including the extra measurements of $\Delta m^2_{ee}$ in Eq.
(\ref{dmee}) and $\Delta m^2_{\mu\mu}$ in Eq. (\ref{dmmu}).

For a reactor neutrino experiment at the medium baseline,
corrections to the mass-squared differences from the terrestrial
matter effect are around $1\%$ and the induced uncertainties are
negligibly small (less than $0.1\%$). On the other hand, in the
muon-neutrino disappearance channel of long-baseline accelerator
neutrino experiments, the matter corrections are suppressed by the
smallness of $\theta_{13}$ and only at the level of $0.2\%$ for the
baselines of several hundreds kilometers (e.g., 295 km for T2K
\cite{T2Kpro} and 735 km
for NOvA \cite{Novapro}). 
Moreover, the different signs in the matter potentials of neutrino
and antineutrino oscillations are also favorable to increase the
discrepancy of different mass-squared differences.

\section{Statistical Analysis}

\par
The 20 kt liquid scintillator detector of Daya Bay II
Experiment~\cite{talk1,talk2,talk3} will be located at equal
baselines of 52 km away from two reactor complexes (36 GW in total).
In this study we use nominal running time of six years, 300
effective days per year, and a detector energy resolution
$3\%/\sqrt{E{\rm (MeV)}}$ as a benchmark. A normal MH is assumed to
be the true one (otherwise mentioned explicitly) while the
conclusion won't be changed for the other assumption. The relevant
oscillation parameters are taken from the latest global analysis
\cite{global1} as $\Delta m^2_{21}=7.54\times10^{-5} {\rm eV}^{-2}$,
$(\Delta m^2_{31}+\Delta m^2_{32})/2=2.43\times10^{-5} {\rm
eV}^{-2}$, $\sin^2\,\theta_{13}=0.024$ and
$\sin^2\theta_{12}=0.307$. The CP-violating phase will be specified
when needed. Finally, the reactor antineutrino flux model from Vogel
{\it et al.}~\cite{Vogel} is adopted in our simulation\footnote{We
have tried both the calculated~\cite{Vogel} and the new
evaluations~\cite{fluxnew1,fluxnew2} of the reactor antineutrino
fluxes. The discrepancy only influences the measurement of
$\theta_{12}$. Both evaluations give consistent results on the MH
determination.}. Because two of the three mass-squared differences
($\Delta m^2_{21}$, $\Delta m^2_{31}$ and $\Delta m^2_{32}$) are
independent, we choose $\Delta m^2_{21}$ and $\Delta m^2_{ee}$
defined in Eq. (\ref{dmee}) as the free parameters in this work.

To obtain the sensitivity of the proposed experiment, we employ the
least squares method and construct a standard $\chi^2$ function as
following:
\begin{equation}
\chi^2_{\text{REA}}=\sum^{N_{\text{bin}}}_{i=1}\frac{[M_{i} -
T_{i}(1+\sum_k \alpha_{ik}\epsilon_{k})]^2}{M_{i}} +
\sum_k\frac{\epsilon^2_{k}}{\sigma^2_k}\,,\label{chiREA}
\end{equation}
where $M_{i}$ is the measured neutrino events in the
$i$-th energy bin, 
$T_{i}$ is the predicted reactor antineutrino flux with
oscillations, $\sigma_k$ is the systematic uncertainty,
$\epsilon_{k}$ is the corresponding pull parameter, and
$\alpha_{ik}$ is the fraction of neutrino event contribution of the
$k$-th pull parameter to the $i$-th energy bin. The considered
systematic uncertainties include the correlated (absolute) reactor
uncertainty ($2\%$), the uncorrelated (relative) reactor uncertainty
($0.8\%$), the flux spectrum uncertainty ($1\%$) and the
detector-related uncertainty ($1\%$). We use 200 equal-size bins for
the incoming neutrino energy between 1.8 MeV and 8.0 MeV.
\begin{figure}
\begin{center}
\begin{tabular}{cc}
\includegraphics*[bb=20 20 290 232, width=0.46\textwidth]{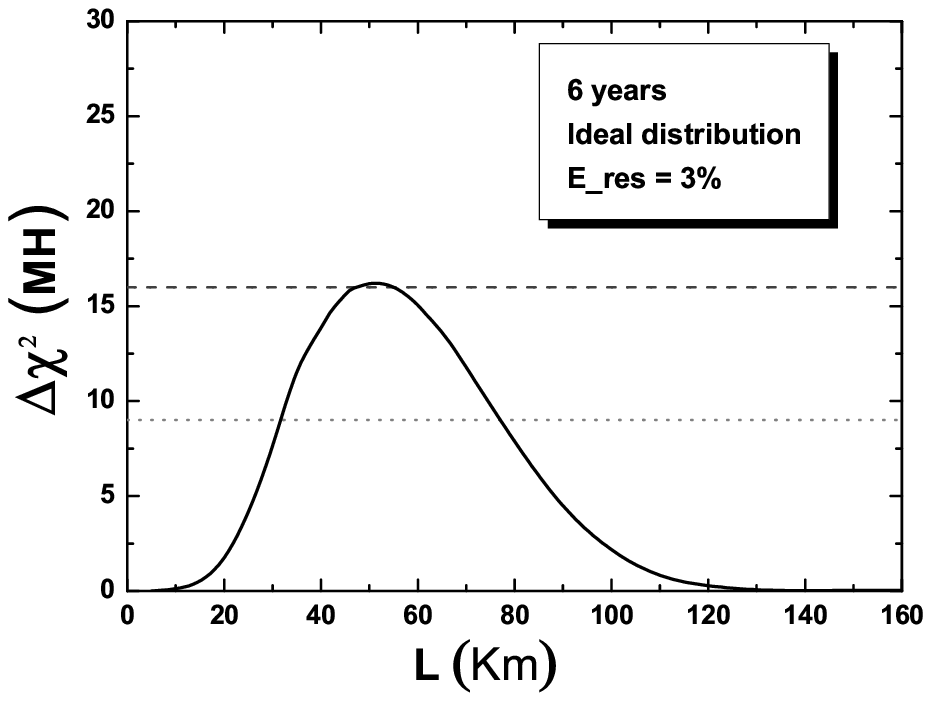}
&
\includegraphics*[bb=20 18 290 230, width=0.46\textwidth]{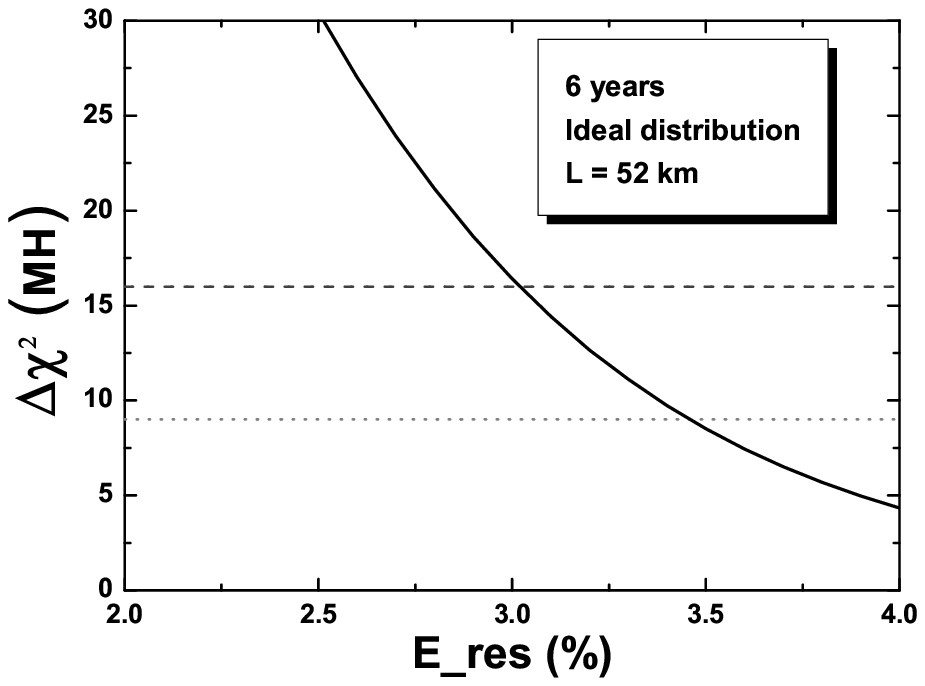}
\end{tabular}
\end{center}
\caption{The MH discrimination ability for the proposed reactor
neutrino experiment as functions of the baseline (left panel) and
the detector energy resolution (right panel) with the method of the
least squares function in Eq. (\ref{chiREA}).}
\end{figure}

We can fit both the
normal MH and inverted MH with the least squares method and take
the difference of the minima as a
measurement of the MH sensitivity.
The discriminator of the
neutrino MH can be defined as
\begin{equation}
\Delta \chi^2_{\text{MH}}=|\chi^2_{\rm min}(\rm N)-\chi^2_{\rm
min}(\rm I)|,
\end{equation}
where the minimization process is implemented for all the relevant
oscillation parameters. Note that two local minima for each MH
[$\chi^2_{\rm min}(\rm N)$ and $\chi^2_{\rm min}(\rm I)$] can be
located at different positions of $|\Delta m^2_{ee}|$. This
particular discriminator is used to obtain the optimal baseline
and to explore the impact of the energy resolution, which are shown
in the left and right panels of Figure 1. Ideally a sensitivity of
$\Delta \chi^2_{\text{MH}}\simeq16$ can be obtained at the baseline around 50 km and
with a detector energy resolution of $3\%$.

\begin{table}
\centering
\begin{tabular}{|c|c|c|c|c|c|c|}\hline\hline
Cores & YJ-C1 & YJ-C2 & YJ-C3 & YJ-C4 & YJ-C5  & YJ-C6 \\
\hline Power (GW) & 2.9 & 2.9 & 2.9 & 2.9 & 2.9 & 2.9 \\ \hline
Baseline(km) & 52.75 & 52.84 & 52.42 & 52.51 & 52.12 & 52.21 \\
\hline\hline
Cores & TS-C1 & TS-C2 & TS-C3 & TS-C4 & DYB  & HZ \\
\hline Power (GW) & 4.6 & 4.6 & 4.6 & 4.6 & 17.4 & 17.4 \\ \hline
Baseline(km) & 52.76 & 52.63 & 52.32 & 52.20 & 215 & 265 \\
\hline
\end{tabular}
\caption{Summary of the power and baseline distribution for the
Yangjiang (YJ) and Taishan (TS) reactor complexes, as well as the remote reactors of Daya Bay (DYB) and
Huizhou (HZ).}
\end{table}

\begin{figure}
\begin{center}
\begin{tabular}{cc}
\includegraphics*[bb=20 16 284 214, width=0.46\textwidth]{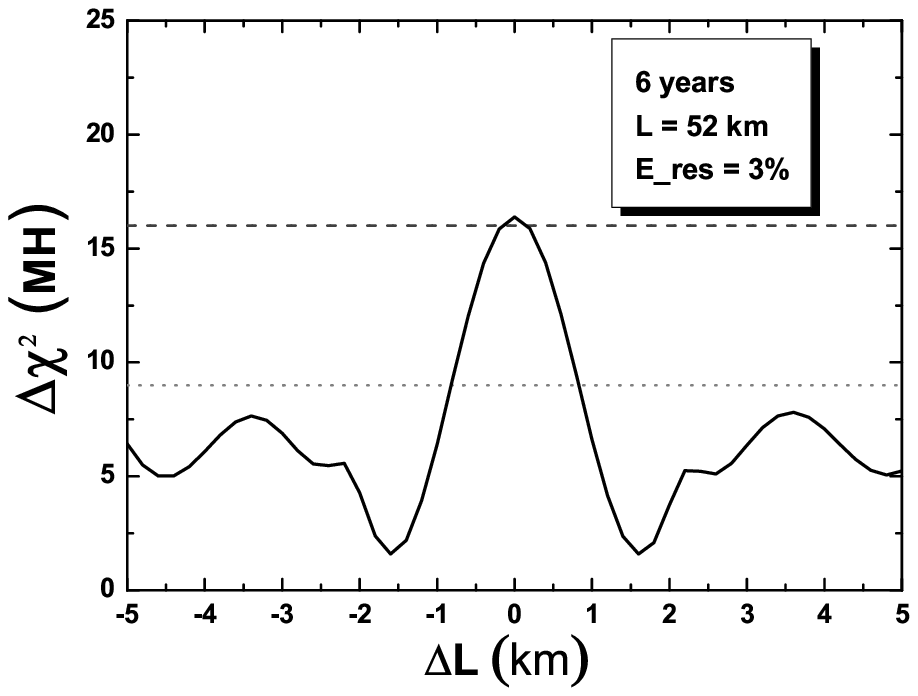}
&
\includegraphics*[bb=26 22 292 222, width=0.46\textwidth]{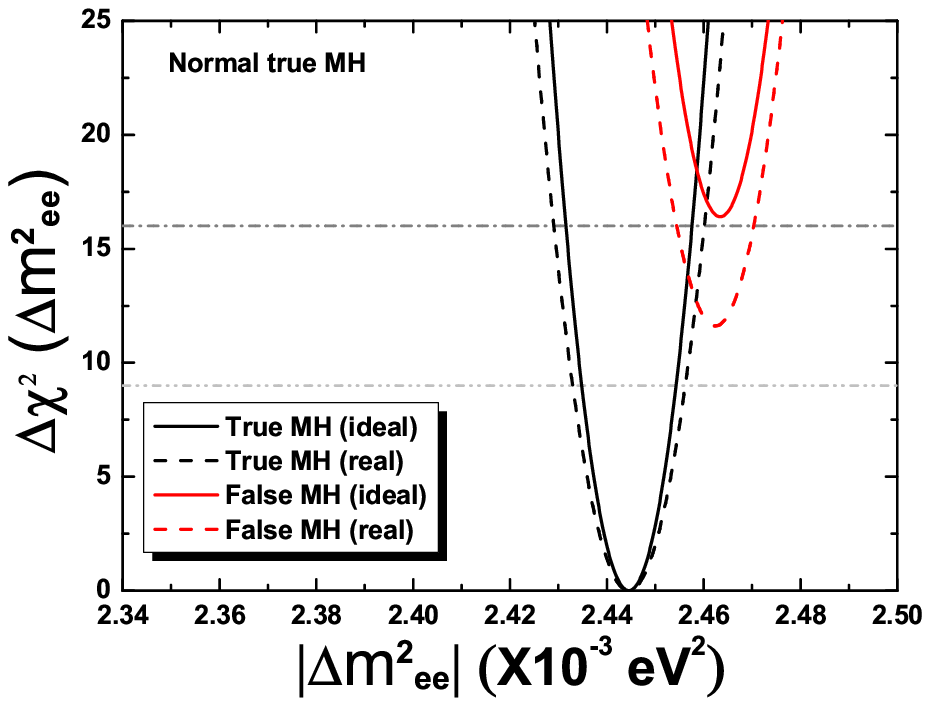}
\end{tabular}
\end{center}
\caption{The variation (left panel) of the MH sensitivity as a
function of the baseline difference of two reactors and the
comparison (right panel) of the MH sensitivity for the ideal and
actual distributions of the reactor cores.}
\end{figure}

The baselines to two reactor complexes should be equal. The impact
of unequal baselines is shown in the left panel of Figure 2, by
keeping the baseline of one reactor unchanged and varying that of
another. A rapid oscillatory behavior is observed and demonstrates
the importance of baseline differences for the reactor cores. To
evaluate the impact from the spacial distribution of individual
cores, we take the actual power and baseline distribution of each
core of the Yangjiang (YJ) and Taishan (TS) nuclear power plant,
shown in Table 1. The remote reactors in the Daya Bay (DYB) and the
possible Huizhou (HZ) power plant are also included. The reduction
of sensitivity due to the actual distribution of reactor cores is
shown in the right panel of Figure 2, which gives a degradation of
$\Delta \chi^2_{\text{MH}}\simeq5$. In all the following studies,
the actual spacial distribution of reactor cores for the Daya Bay II
Experiment is taken into account.

\begin{figure}
\begin{center}
\begin{tabular}{cc}
\includegraphics*[bb=20 16 284 214, width=0.5\textwidth]{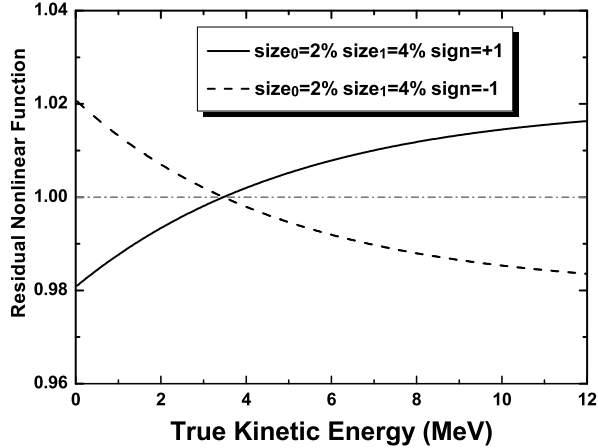}
\end{tabular}
\end{center}
\caption{Two classes of typical examples for the residual non-linear
functions in our simulation.}
\end{figure}

\section{Energy Non-Linearity Effect}

The detector energy response is also crucial for Daya Bay II since a
precise energy spectrum of reactor neutrinos is required.
Assuming the energy non-linearity correction is imperfect, we study its impact to the sensitivity by
including in our simulation a residual non-linearity between the measured and expected neutrino spectra. Assume the detector energy non-linearity has the
form as
\begin{eqnarray}
\frac{E_{\rm rec}}{E_{\rm true}}= \frac{1+p_0}{1+p_1\exp(-p_2E_{\rm
true})}\,, \label{residual}
\end{eqnarray}
where ${E_{\rm rec}}$ and ${E_{\rm true}}$ are the reconstructed and
true kinematic energy of the positron from the inverse beta decay,
respectively. The parameter of $p_0$, $p_1$ and $p_2$ describe the
shape and magnitude of the non-linear functions. We assume that,
after the non-linearity correction, the residual non-linearity in
measured energy spectrum also has the same function form. The
conclusion won't be changed for other residual non-linearity
assumptions, because we will use a quadratic function in the
predicted spectrum, different from the measured spectrum. We fix
$p_2=0.2\,/\text{MeV}$ and vary $p_0$ and $p_1$ as
\begin{eqnarray}
p_0=\text{sign}\times \text{size}_0\quad\quad  \text{and}\quad\quad
p_1=\text{sign}\times\text{size}_1\,,
\end{eqnarray}
where $\text{sign}=\pm1$ determines the slope, $\text{size}_0$ and
$\text{size}_1$ can be a few percent to indicate the magnitudes of
the residual non-linearity. Two typical examples with
$\text{sign}=\pm1$ are shown in Figure 3.

By including the residual non-linearity in $M_i$ of Eq.
(\ref{chiREA}), we obtain in Figure 4 updates on the distributions
of the $\Delta \chi^2$ function for both true and false neutrino MH,
where normal (inverted) MH is assumed to be the true one in upper
(lower) panels. Different classes of the non-linear functions may
induce different effects on the MH determination. Comparing to that
without non-linearity effects as shown in the right panel of Figure
2, the non-linearity with positive sign ($\text{sign}=+1$) will
increase the discrepancy between two neutrino MH scenarios for the
normal true MH, but decrease the discrepancy for the inverted true
MH. The non-linear functions with negative sign ($\text{sign}=-1$)
have opposite effect. Only when the size of non-linearity is as
small as $0.5\%$ this effect can be ignored.
\begin{figure}
\begin{center}
\begin{tabular}{cc}
\includegraphics*[bb=25 20 295 228, width=0.46\textwidth]{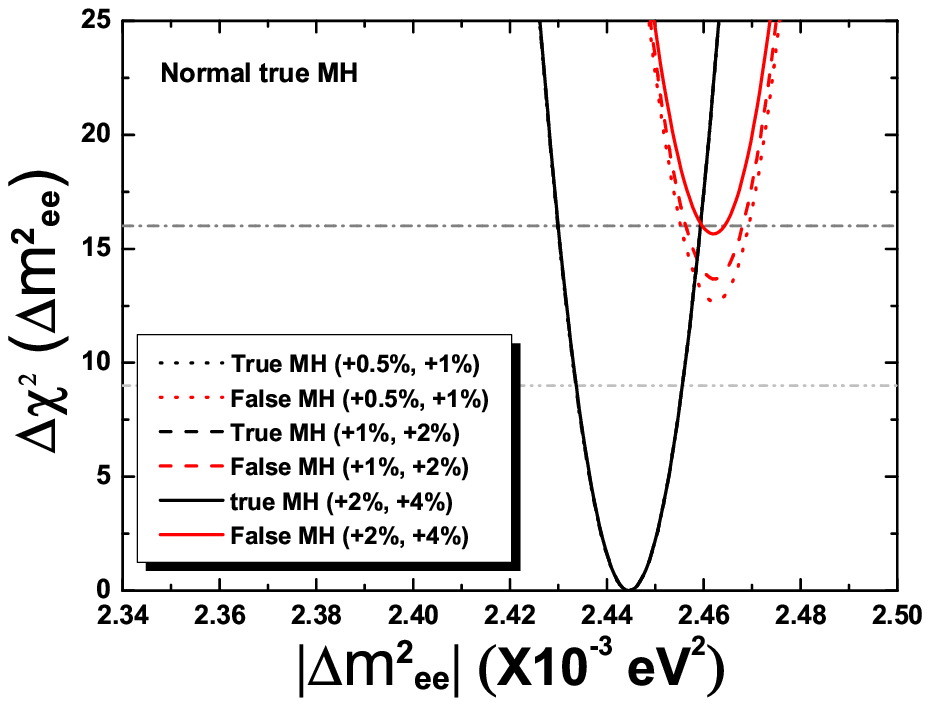}
&
\includegraphics*[bb=25 20 295 240, width=0.46\textwidth]{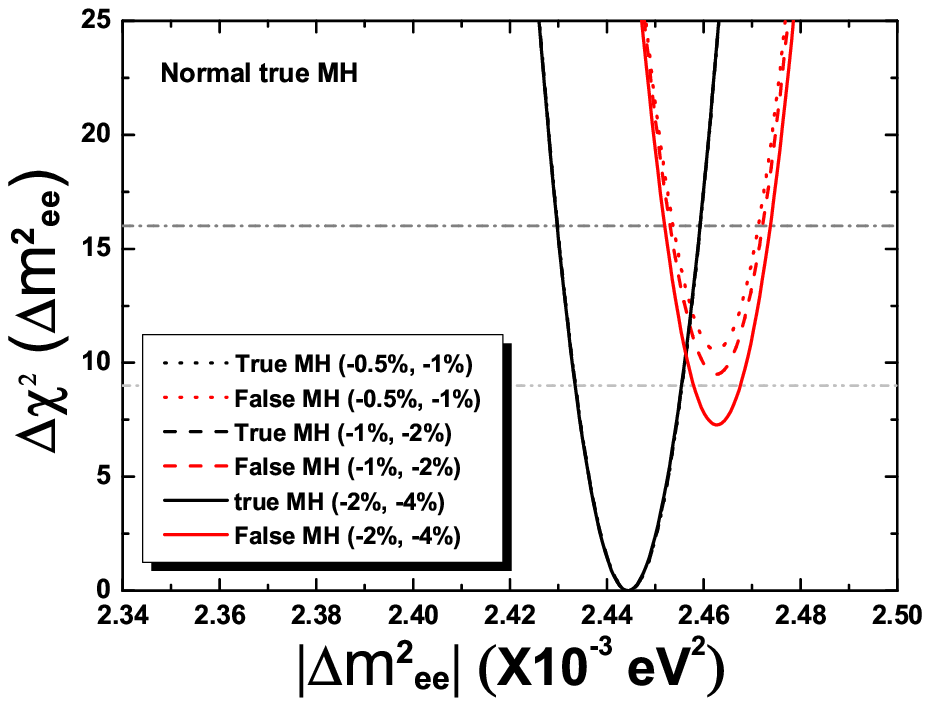}
\\
\includegraphics*[bb=25 20 295 228, width=0.46\textwidth]{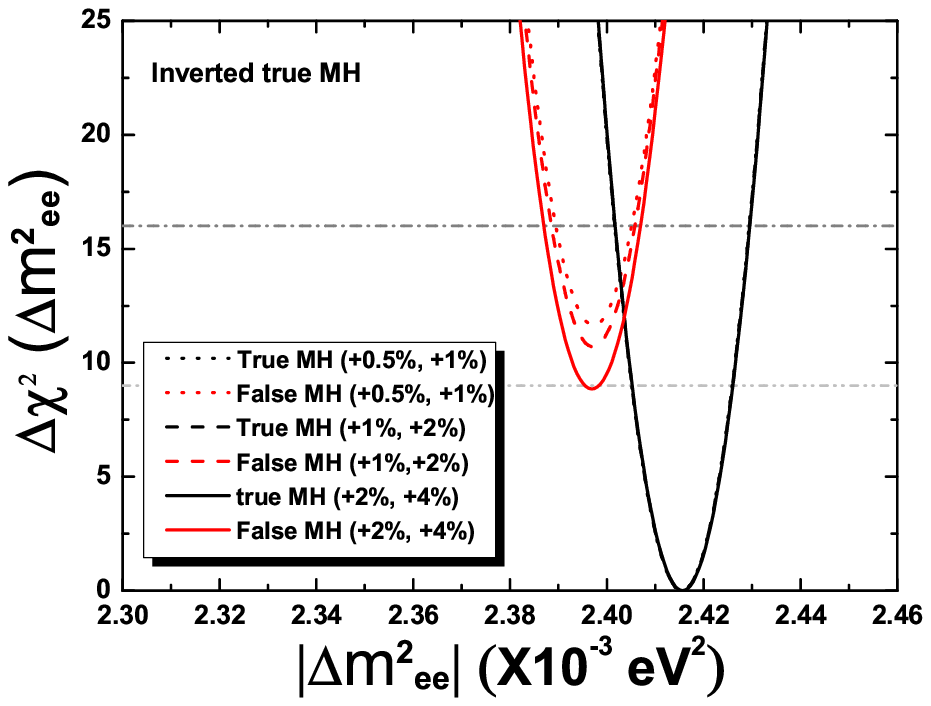}
&
\includegraphics*[bb=25 20 295 240, width=0.46\textwidth]{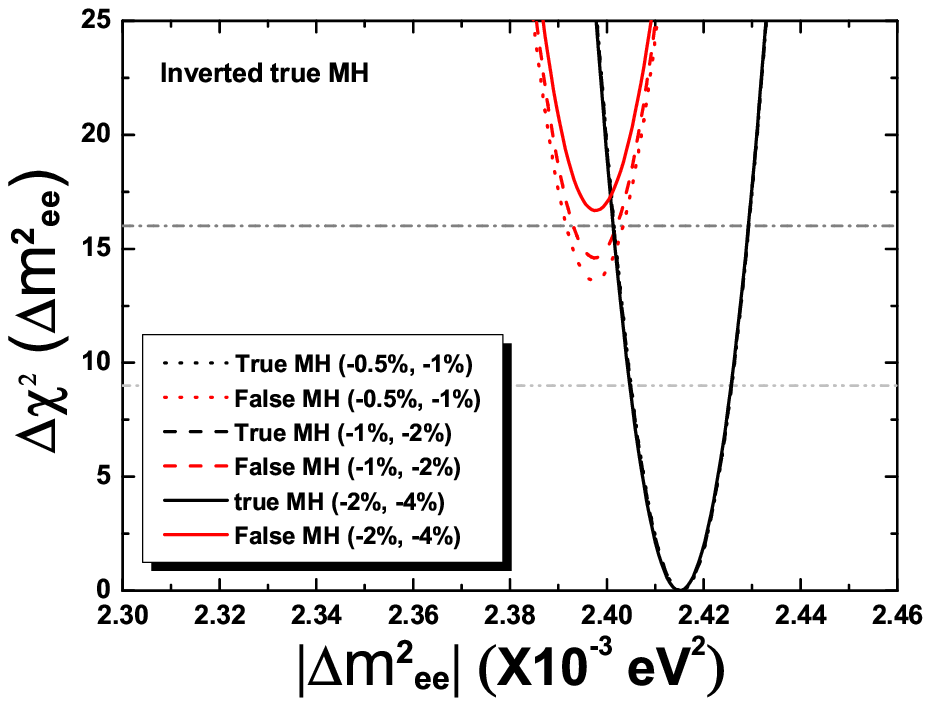}
\end{tabular}
\end{center}
\caption{Effects of two classes of energy non-linearity models in
the determination of the neutrino MH without the self-calibration in
fitting. The normal (inverted) MH is assumed to be the true one in
the upper (lower) panels. The sign and size of the non-linear
parameters in the form of ($p_0$, $p_1$) are indicated in the
legend.}
\end{figure}

\begin{figure}
\begin{center}
\begin{tabular}{cc}
\includegraphics*[bb=25 20 295 228, width=0.46\textwidth]{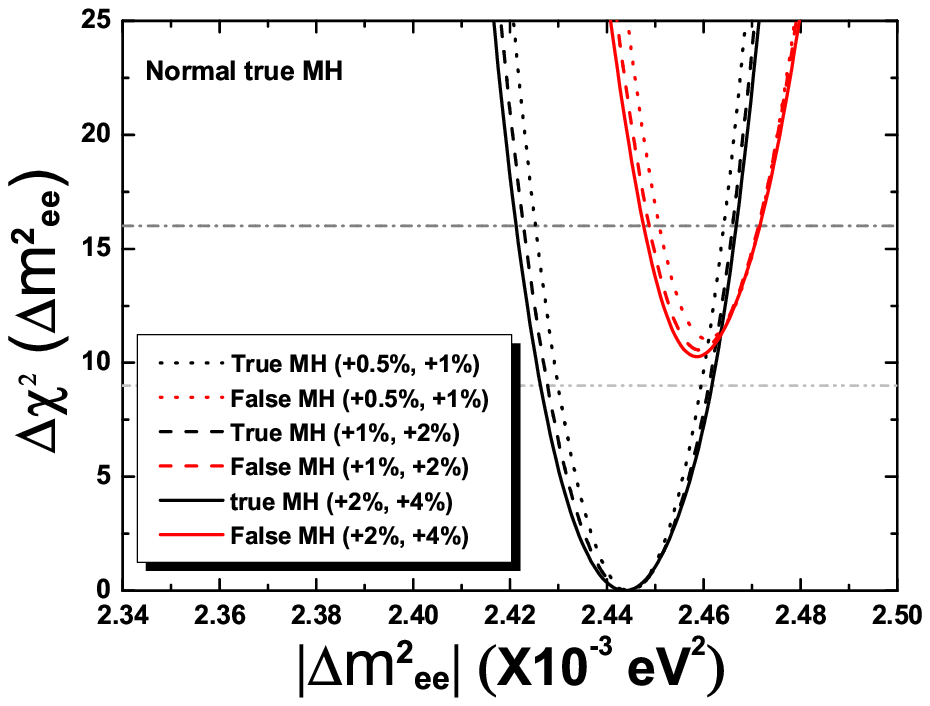}
&
\includegraphics*[bb=25 20 295 240, width=0.46\textwidth]{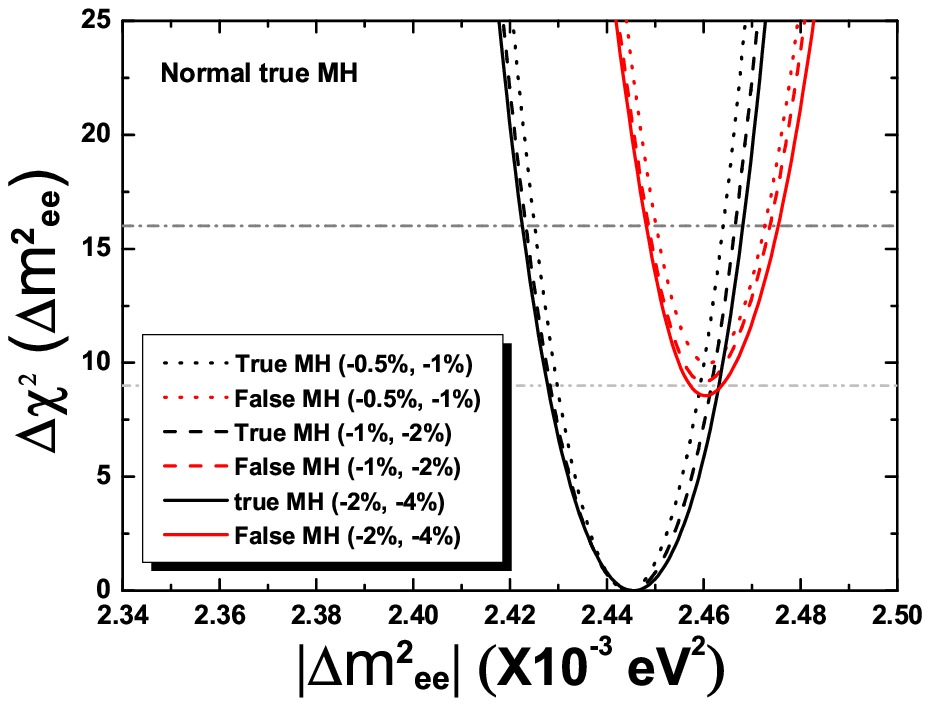}
\\
\includegraphics*[bb=25 20 295 228, width=0.46\textwidth]{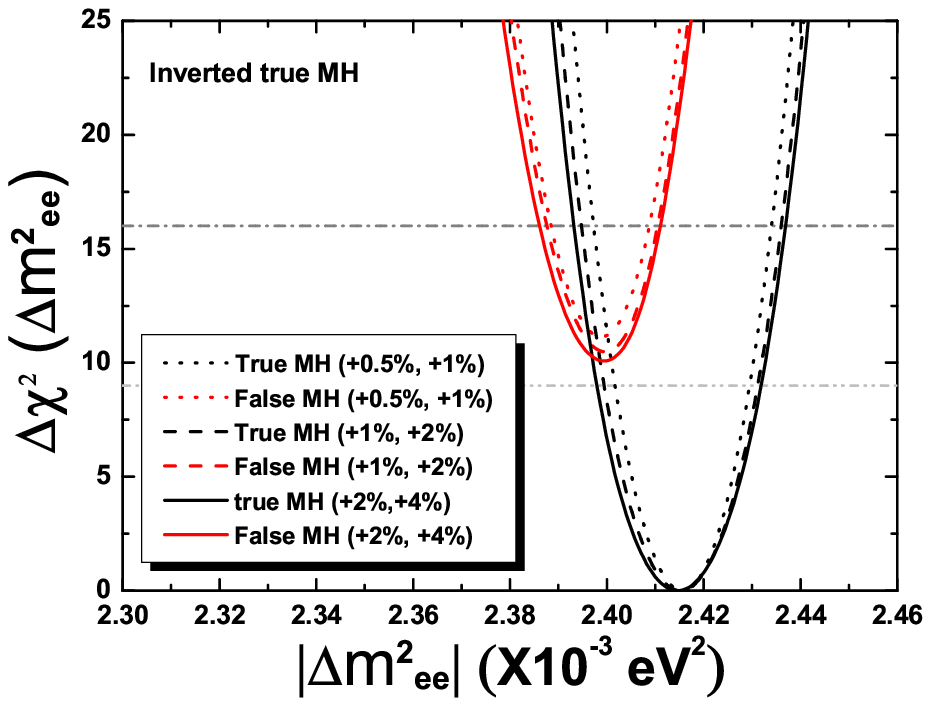}
&
\includegraphics*[bb=25 20 295 240, width=0.46\textwidth]{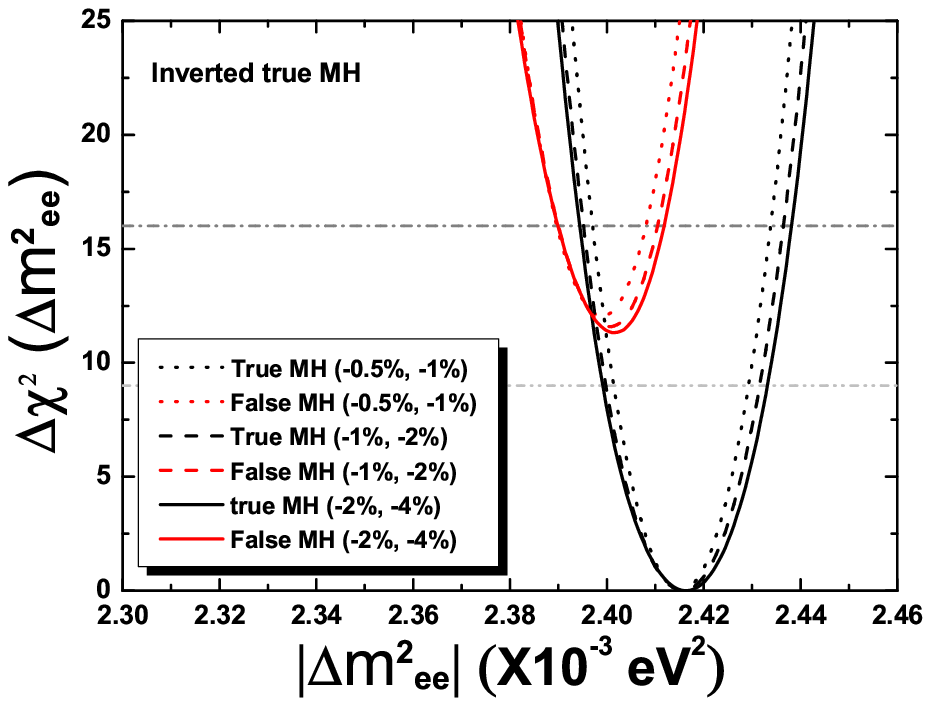}
\end{tabular}
\end{center}
\caption{Effects of two classes of energy non-linearity models in
the determination of the neutrino MH with the self-calibration in
fitting. The normal (inverted) MH is assumed to be the true one in
the upper (lower) panels. The sign and size of the non-linear
parameters in the form of ($p_0$, $p_1$) are indicated in the
legend.}
\end{figure}

For the reactor antineutrino experiment at medium baseline ($\sim50$
km), we can observe multiple peaks of the $\Delta m^2_{ee}$ induced
oscillation. Each peak position carries the information of $\Delta
m^2_{ee}$. This redundancy can be used to evaluate the energy scale
at different energies. Therefore, we can measure to some extent the
energy non-linearity by the spectrum itself. We call this effect as
the self-calibration of the spectrum. To illustrate, we consider a
test quadratic non-linear function in the fitting process,
\begin{eqnarray}
\frac{E_{\rm rec}}{E_{\rm true}}\simeq 1+q_0 +q_1 E_{\rm true} +
q_{2} E^2_{\rm true}\,,
\end{eqnarray}
where the central values of three parameters are arbitrary, but the
uncertainties are assumed to be the same size as the function in Eq.
(\ref{residual}). We can define the new least-square function by
using the test non-linearity function in $T_i$ and including the
corresponding pull terms $\chi^2_{\text{NL}}=
\sum^{2}_{i=0}{q^2_{i}}/{(\delta q_{i})^2}\,$ and derive the MH
sensitivity taking into account the non-linearity and self-calibration
effects. Considering different sign and size of nonlinearity defined
in Eq. (13), we illustrate the $\Delta \chi^2$ function with the
self-calibration in fitting in Figure 5, where the normal (inverted) MH is
assumed to be the true one in the upper (lower) panels. First we can observe that the
degeneracy ambiguity induced by the non-linearity effect can be
removed by fitting the parameters of the test non-linearity function
and both classes of non-linear functions give the consistent
sensitivity of MH determination (see the left and right panels of
Figure 4 and 5). Tiny differences can be noticed for different size
of the non-linearity because the test quadratic function cannot
describe the true residual non-linearity accurately. Second, the width of the $\Delta
\chi^2$ functions in Figure 5 is broadened compared to Figure 4.
This is because additional uncertainties from the non-linearity
parameters are introduced, which can be translated to the
uncertainty of the neutrino spectrum and finally to the accuracy of
the oscillation parameters.

\section{Improvement with External Measurements}

Taking into account different definitions of $\Delta
m^2_{\alpha\alpha}$ in different oscillation scenarios, precise
measurements of $\Delta m^2_{ee}$ and $\Delta m^2_{\mu\mu}$ in Eqs.
(\ref{dmee}) and (\ref{dmmu}) can provide the additional MH
sensitivity in Daya Bay II.

To incorporate the contributions of $\Delta m^2_{\mu\mu}$ from
long-baseline accelerator neutrino experiments and $\Delta m^2_{ee}$
from short-baseline reactor neutrino experiments, we define the
following pull $\chi^2$ function
\begin{equation}
\chi^2_{\text{pull}}= \frac{(|\Delta m^2_{\mu\mu}|-|\overline{\Delta
m^2_{\mu\mu}}|)^2}{\sigma^2(\Delta m^2_{\mu\mu})} + \frac{(|\Delta
m^2_{ee}|-|\overline{\Delta m^2_{ee}}|)^2}{\sigma^2(\Delta
m^2_{ee})}\,,\label{chipull}
\end{equation}
where $\overline{\Delta m^2_{\mu\mu}}$ ($\overline{\Delta
m^2_{ee}}$) and $\sigma(\Delta m^2_{\mu\mu})$ ($\sigma(\Delta
m^2_{ee})$) are the central value and $1\sigma$ uncertainty of the
measurement respectively. The combined $\chi^2$ function is defined
as
\begin{equation}
\chi^2_{\text{ALL}}=\chi^2_{\text{REA}}+\chi^2_{\text{pull}}\,.\label{chiALL}
\end{equation}
As mentioned in the previous Section, we choose $\Delta m^2_{21}$
and $\Delta m^2_{ee}$ defined in Eq. (\ref{dmee}) as the free
parameters. The values of $\Delta m^2_{\mu\mu}$ can be calculated by
the relations in Eqs. (8) and (9) by assuming different choices of
the MH.

In general we need to consider the uncertainties of other
oscillation parameters, but the CP-violating phase $\delta$ is
almost unconstrained and we can absorb the uncertainties of other
parameters in that of the phase $\delta$ and consider the whole
range of CP-violating phase from $0^\circ$ to $360^\circ$. Until
2020, the most accurate measurement of $\Delta m^2_{ee}$ may come
from the Daya Bay experiment, where an accuracy of $4\%$
\cite{DYBfuture} can be achieved after a 3-year running of full
operation. Numerical analysis demonstrates that the measurement of
$\Delta m^2_{ee}$ at this level is negligible in the $\chi^2$
function in Eq. (\ref{chipull}). Therefore, we focus on the effect
of $\Delta m^2_{\mu\mu}$ in this work.

As shown in Eq. (\ref{dmemu}), because the relative size of $\Delta
m^2_{ee}$ and $\Delta m^2_{\mu\mu}$ is different for the normal and
inverted neutrino MH, the extra pull function $\chi^2_{\text{pull}}$
in Eq. (\ref{chipull}) can give a non-zero contribution to the
discriminator $\Delta \chi^2_{\text{MH}}$ at the magnitude of
\begin{equation}
\chi^2_{\text{pull}}(\rm MH)\sim\frac{[2\times\Delta
m^2_{21}(\cos2\theta_{12}-\sin2\theta_{12}\sin\theta_{13}\tan\theta_{23}\cos\delta)]^2}{\sigma^2(\Delta
m^2_{\mu\mu})}\,,\label{chipull2}
\end{equation}
and accordingly improve the neutrino MH sensitivity for the reactor
neutrino experiment at medium baseline.

\begin{figure}
\begin{center}
\begin{tabular}{cc}
\includegraphics*[bb=25 20 295 228, width=0.46\textwidth]{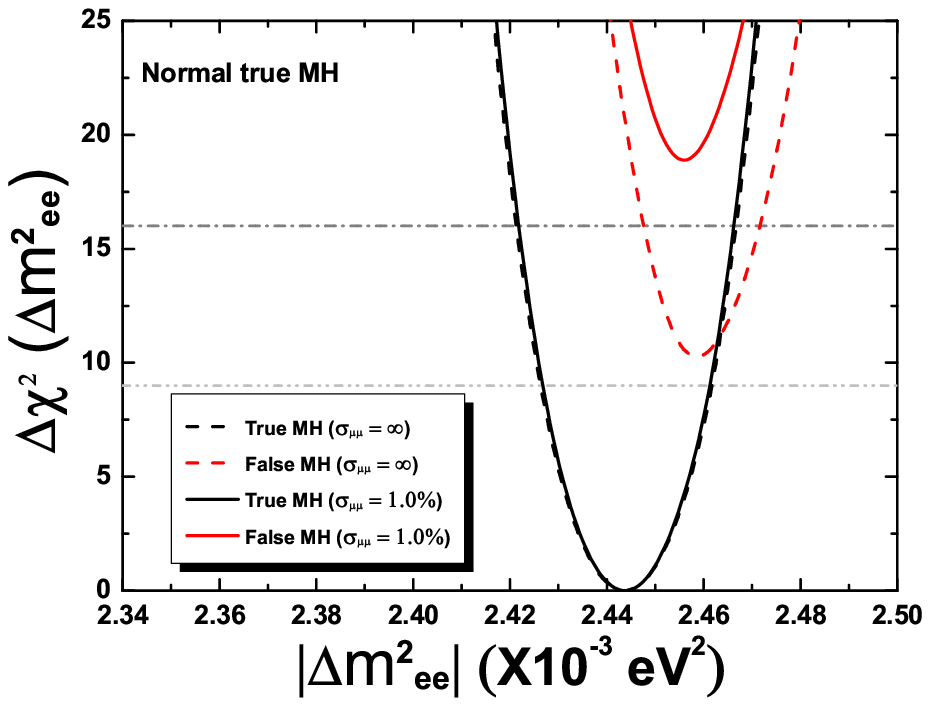}
&
\includegraphics*[bb=25 20 295 228, width=0.46\textwidth]{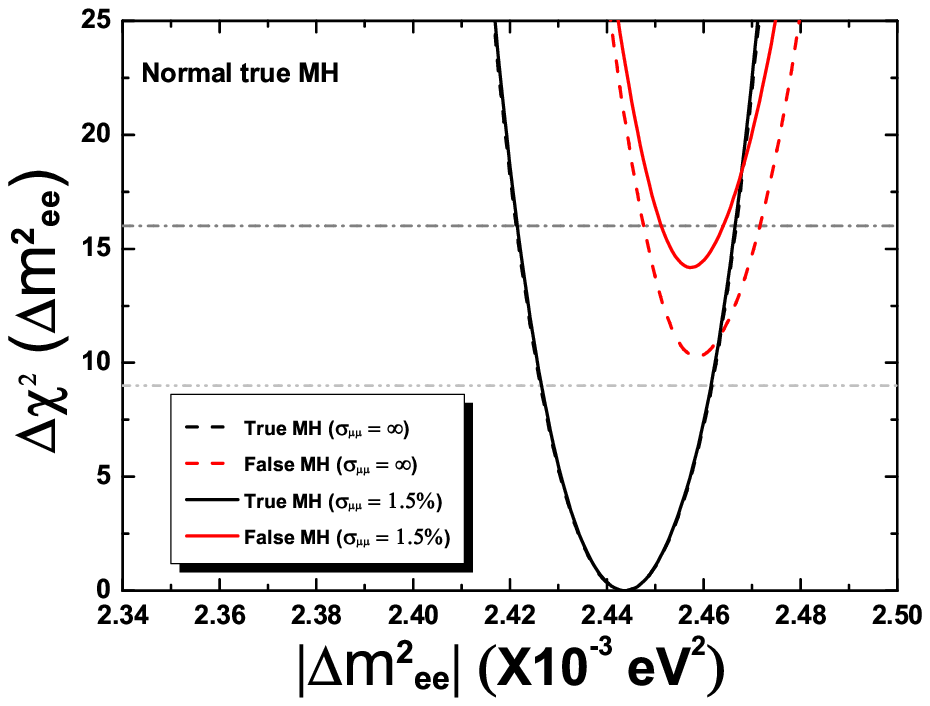}
\end{tabular}
\end{center}
\caption{the reactor-only (dashed) and combined (solid)
distributions of the $\Delta\chi^2$ function in Eq. (\ref{chiREA})
and Eq. (\ref{chiALL}), where a $1\%$ (left panel) or $1.5\%$ (right
panel) relative error of $\Delta m^2_{\mu\mu}$ is assumed and the
CP-violating phase ($\delta$) is assigned to be $90^\circ/270^\circ$
($\cos\delta=0$) for illustration. The black and red lines are for
the true (normal) and false (inverted) neutrino MH, respectively.
The non-linearity in Eq. (\ref{residual}) is assigned with
$\text{sign}=+ 1$, $\text{size}_0=2\%$ and $\text{size}_1=4\%$.}
\end{figure}

To illustrate the effect of the external $\Delta m^2_{\mu\mu}$
measurement, we first fix the non-linearity with  $\text{sign}=+ 1$,
$\text{size}_0=2\%$ and $\text{size}_1=4\%$, choose $\delta$ to be
$90^\circ/270^\circ$ ($\cos\delta=0$) and give the separated and
combined distributions of the $\chi^2$ functions in Eqs.
(\ref{chiREA}) and (\ref{chiALL}) in Figure 6, where a $1\%$ (left
panel) or $1.5\%$ (right panel) relative error of $\Delta
m^2_{\mu\mu}$ is assumed. The black and red lines are for the true
(normal) and false (inverted) neutrino MH, respectively. The dashed
and solid lines are for the reactor-only [in Eq. (\ref{chiREA})] and
combined distributions. We can get a value of $\Delta
\chi^2_{\text{MH}}\simeq(10\div11)$ for the reactor-only analysis in
the least-squares method. As for the contribution from the external
$\Delta m^2_{\mu\mu}$ measurement, it is almost negligible if we
choose the true MH in the fitting program. However, if the fitting
MH is different from the true one, the central value of $\Delta
m^2_{ee}$ in the $\chi^2_{\text{pull}}$ function will change by two
units of the difference in Eq. (\ref{dmemu}), which accordingly
results in a significant contribution to the combined $\chi^2$
function. Finally we can achieve $\Delta \chi^2_{\text{MH}}\simeq19$
and $\Delta \chi^2_{\text{MH}}\simeq14$ for the $1\%$ and $1.5\%$
relative errors of the $\Delta m^2_{\mu\mu}$ measurement.
\begin{figure}
\begin{center}
\begin{tabular}{c}
\includegraphics*[bb=20 20 290 230, width=0.6\textwidth]{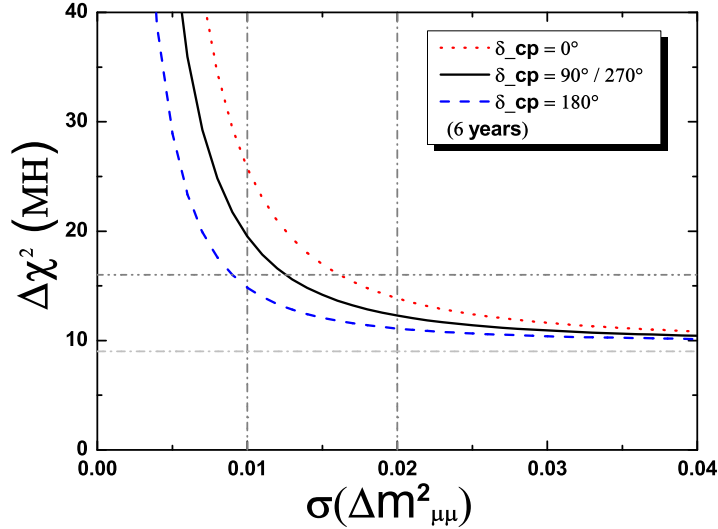}

\end{tabular}
\end{center}
\caption{The $\Delta \chi^2_{\text{MH}}$ dependence on different
input errors of $\Delta m^2_{\mu\mu}$ is illustrated. The blue
dashed, black solid and red dotted lines stands for different CP
values ($\delta=0^\circ$, $\delta=90^\circ/270^\circ$ and
$\delta=180^\circ$ respectively).}
\end{figure}

Next we can discuss the ambiguity of the unknown CP-violating phase
$\delta$ and evolution of the MH sensitivity with respect to changes
of the $\Delta m^2_{\mu\mu}$ error. The $\Delta \chi^2_{\text{MH}}$
dependence on different input errors is shown in Figure 7, where the
blue, black and red lines stands for different values of the
CP-violating phase ($\delta=0^\circ$, $\delta=90^\circ/270^\circ$
and $\delta=180^\circ$ respectively). In Figure 7, we can notice
that the improvement are obvious for an external $\Delta
m^2_{\mu\mu}$ measurement better than $2\%$ and becomes significant
if we can get to the $1\%$ level. For the effect of the CP-violating
phase, it is most favorable for the value close to $180^\circ$. The
cases of maximal CP violation are in the middle region which are
just the cases discussed in Figure 6. The ambiguity of the
CP-violating phase can induce an uncertainty of $\Delta
\chi^2_{\text{MH}}\simeq2\,(4)$ at $\sigma(\Delta
m^2_{\mu\mu})/|\Delta m^2_{\mu\mu}|\simeq 1.5\%\,(1\%)$. The effect
of the external $\Delta m^2_{\mu\mu}$ measurement can also be viewed
as a probe of the CP-violating phase. If the improvement is much
better than the discussion in Figure 6, a preference of $\delta$
close to $180^\circ$ can be achieved. Otherwise, we may get a nearly
vanishing CP-violating phase if the situation is totally opposite.

Current best measurement for $\Delta
m^2_{\mu\mu}$ from the MINOS
experiment \cite{MINOS2} gives an error of $4\%$. Two new
experiments T2K \cite{T2Kpro} and NOvA \cite{Novapro} are in
operation or construction and each of them can reach $1.5\%$ by 2020
after finishing of their nominal running plans (5 years of $\nu$
mode at 750 kW for T2K and 3 years of $\nu$ mode plus 3 years of
$\bar\nu$ mode at 700 kW for NOvA). If these experiments could
extend to another 5-year running, it might be possible to obtain the
precision of $1\%$ which will be useful for the measurement of the
precision reactor neutrino experiment.

\section{Conclusion}
In this work, we have discussed the determination of the neutrino MH
using the Daya Bay II reactor neutrino experiment at a medium
baseline around 50 km away from reactors. Precision measurements of
the reactor antineutrino spectrum can probe the interference effect
of two fast oscillation modes (i.e., oscillations induced by $\Delta
m^2_{31}$ and $\Delta m^2_{32}$) and sensitive to the neutrino MH.
The corresponding sensitivity depends strongly on the size of
$\theta_{13}$, the energy resolution, the baseline differences and
energy response functions. Moreover, the MH sensitivity can be
improved by including a measurement of the effective mass-squared
difference in the long-baseline muon-neutrino disappearance
experiment due to flavor dependence of the effective mass-squared
differences.

We have calculated the MH sensitivity taking into account the real
spatial distribution of reactor complexes, and demonstrated that the
residual energy non-linearity of the liquid scintillator detector
has limited impacts on the sensitivity due to the self-calibration
of small oscillation peaks. We numerically calculated the
sensitivity by assuming two typical classes of energy non-linearity
functions ($2\%$) and discussed the improvement with the external
$\Delta m^2_{\mu\mu}$ measurement quantitatively. To conclude, the
Daya Bay II Experiment could determine the mass hierarchy
unambiguously with a confidence level of $\Delta
\chi^2_{\text{MH}}\sim14$ ($3.7\,\sigma$) or $\Delta
\chi^2_{\text{MH}}\sim19$ ($4.4\,\sigma$) in 6 years, for the
$\Delta m^2_{\mu\mu}$ uncertainty of 1.5\% or 1\%.



\end{document}